%%%% 
\input amstex
%\input epsf
%\input preamble
%\newsymbol\boxtimes 1202

%letters

\def\b1{\text{\bf 1}}

\def\#{\,\check{}}

%symbols

% space between paragraphs 
%\parskip=6pt

\documentstyle{amsppt}

\NoBlackBoxes

\topmatter
\title     An analog of the Feynman-Kac formula for Dirac's electron in electromagnetic field and the correspondence principle   \endtitle
\author A.~A.~Beilinson  \endauthor
\leftheadtext{A.~A.~Beilinson}

\address 
\endaddress

 \keywords  generalized functions, generalized  functionals, the support of a functional, the Foldy-Wouthuysen presentation, Zitterbewegung \endkeywords

\email  \endemail

\thanks \endthanks

%  Math Subject Classifications 
%\subjclassyear{2000}
%\subjclass  \endsubjclass

\abstract The article describes a relation between the fundamental solutions of Dirac's equations for free electron and electron in given electromagnetic field viewed as functionals on bump functions.

We explain how   classical relativistic mechanics of a charged particle in given electromagnetic field arises from quantum mechanics of Dirac's electron in that field.

   \endabstract

\endtopmatter

\document
%\magnification=1100

%\nologo

\head  Introduction \endhead

We construct a fundamental solution of Dirac's equations for an electron in given exterior electromagnetic field  as a functional on bump functions via   averaging of the exponent of the classical action of a massless charge in that field by a generalized complex matrix-valued 
functional of Cauchy-Dirac $\Bbb D^m_t \{ dx_\tau \}$ (a generalized function of infinitely many variables) on bump functionals $\varphi 
\{ dx_\tau \}\in K^{(\infty )}$ whose support is compact in the uniform convergence topology with velocities that lie in the Hilbert space. This
generalized  
functional of Cauchy-Dirac that corresponds to the retarded Green's function $D^m_t (x)$ for the free Dirac electron was constructed in
\cite{10}. 

Therefore our approach to the problem is quite similar to the traditional method of taking in account an exterior field viewing it either as a gauge field (see \cite{8}) or using its averaging over the states of a free particle (see \cite{3}).

\medskip
We use E.B.Dynkin's scheme, as was used in \cite{9} to deduce the Feynman-Kac formula in Euclidean non-relativistic quantum mechanics, demanding  the action functional of a charge in given electromagnetic field to be continuous in the uniform convergence topology.

\medskip

For a short exposition of the results see \cite{12}.

\head 1. An analog of the Feynman-Kac formula    \endhead

Recall that the Feynman-Kac formula in non-relativistic quantum mechanics is an expression of a fundamental solution of Schr\"odinger's equation $i{{\partial}\over{\partial t}} \Psi =-{{1}\over{4}}{{\partial^2}\over{\partial x^2}}\Psi + V_t (x)\Psi$ as   Feynman's ``path integral" (see \cite{3}) $$\Psi_t (x_0 ,x_t )=\int_{F\, x(0)=x_0 ,x(t)=x_t} \exp (i\int_0^t (\dot x_\tau^2 -V_\tau (x_\tau ))d\tau )\Pi_{\tau =0}^t  {{dx_\tau}\over{\sqrt{i\pi d\tau}}}  \tag 1$$ and an expression of a fundamental solution of Bloch's equation 
${{\partial}\over{\partial t}} W ={{1}\over{4}}{{\partial^2}\over{\partial x^2}}W - V_t (x)W$ as Wiener's functional integral (see \cite{6}) 
$$W_t (x_0 ,x_t )=\int_{C(0,t)\, x(0)=x_0 ,x(t)=x_t} \exp (-\int_0^t (\dot x_\tau^2 +V_\tau (x_\tau ))d\tau )\Pi_{\tau =0}^t  {{d x_\tau}\over{\sqrt{\pi d\tau}}} $$ (Euclidean quantum mechanics).

Constructions of that kind permit to insert an exterior field in the Hamilton operator via the functional integral defined
by the Hamilton  operator of a free particle. Here the last formula presents the Wiener functional integral which has precise mathematical definition (the Wiener measure is $\sigma$-additive and is supported on a 
set of trajectories which satisfy the H\"older-Lipschitz condition $|x_\tau - x_\sigma |\le C|\tau -\sigma |^{1/2 - \epsilon}$, see e.g., \cite{6}).
 And Feynman's path integral, though being attractive and natural from physicist's viewpoint, is not well defined and hence is of merely heuristic importance.

\medskip

Article \cite{10} studied  coordinate presentation of the fundamental solution $D^m_t (x)$ of Dirac's free electron equation (see \cite{7}, \cite{8}) $$\gamma^0 {{\partial}\over{\partial t}} + (\gamma ,\nabla )+im =0 \tag 2$$ as a functional of space variables on bump functions $K^{(3)}$; here time $t\ge 0$ was a parameter. It was shown that 
$D^m_t (x)$ yields a generalized complex matrix-valued functional (the Cauchy-Dirac quantum process)
$\Bbb D^m_t \{ dx_\tau \}=\Pi^t_{\tau =0} D^m_{ d\tau} (dx_\tau )$ on
the space of bump functionals $\varphi \{ dx_\tau \}
=\varphi (\ldots ,dx_\tau ,\ldots )\in K^{(\infty )}$  whose  support lies in continuous trajectories
 (so the datum of translations $(\ldots ,dx_\tau ,\ldots )$ forms a concrete trajectory   $x_\tau \in   \{ x_\tau \}$)
that are compact in the uniform convergence topology and having velocity ${{dx_\tau}\over{d\tau}}\in L_2 (0,t)$.

\medskip

Therefore, in particular, the fundamental solution $D^m_t (x)$ of Dirac's equation (1) viewed as a functional
on bump functions $\varphi (x)\in K^{(3)}$ (see \cite{1}), can be written as ``path integral" $$\int 
D^m_t (x)\varphi (x)dx=\int_{\{x_\tau \} } (\Pi^t_{\tau =0} D^m_{ d\tau} (dx_\tau ))\varphi (\int^t_0 dx_\tau )
\Pi^t_{\tau =0} dx_\tau 
\tag 3$$ or, in short, $D^m_t (x)=\Pi^t_{\tau =0} *D^m_{ d\tau} (dx_\tau )$ where $*$ is the convolution of  functionals (see \cite{1}, \cite{10}).

\medskip

Our aim is to construct the fundamental solution $D^{me}_t (x)$ of Dirac's equation for an electron in a given external electromagnetic field (see \cite{7}, \cite{8}) $$
i {{\partial}\over{\partial t}}   =(-i\gamma^0    (\gamma ,\nabla )+m\gamma^0 )  +e( -\gamma^0    (\gamma ,A)+A_0 )    . \tag 4$$ Here $\{ A_{0t}(x),A_t (x)\}$ is 4-potential of the field (see \cite{5}).

\medskip

To that end let us consider a matrix-valued functional $$\exp (ie\int^t_0 (\gamma^0    (\gamma ,A_\tau (x_\tau ))- A_{0\tau}(x_\tau ))d\tau )$$ 
and let us assume that the 4-potential $\{ A_0 ,A\}$ is such that the this functional is continuous for the uniform convergence topology. We understand this formula (as well as formulas (5)--(8)) as a chronologically ordered product \cite{8}.

Consider  $$\int_{\{x_\tau \} } (\Pi^t_{\tau =0} D^m_{ d\tau} (dx_\tau ) \exp (ie  (\gamma^0    (\gamma ,A_\tau (x_\tau ))- A_{0\tau}(x_\tau ))d\tau ))\varphi \{ dx_\tau \} \Pi^t_{\tau =0}dx_\tau . \tag 5$$ Here the functional integral (5) is well defined due to compactness of the support  of the complex matrix-valued measure $\Bbb D_t^m 
\{ dx_\tau \}$, and it can be approximated by the corresponding finite-dimensional integral. 

Notice that (5) is a convolution and is similar to the averaging by  $\Bbb D^m_t \{dx_\tau \}=\Pi^t_{\tau =0} D^m_{d\tau}(dx_\tau ).$

Due to its structure functional integral (5) satisfies the Chapman-Kolmogorov equation hence it equals the retarded Green's function (the fundamental solution) of certain equation. We find this equation constructing the infinitesimal operator of the Green's function.

To that end, following E.B.Dynkin's work \cite{9}, notice that $$\exp (ie \int^t_s  (\gamma^0    (\gamma ,A_\tau  )- A_{0\tau} )d\tau )= \tag 6$$ $$= I -    ie \int^t_s  (\gamma^0    (\gamma ,A_\tau  )- A_{0\tau} ) \exp ( ie \int^t_\tau  (\gamma^0    (\gamma ,A_\nu )- A_{0\nu} )d\nu )d\tau     $$ which follows due to absolute continuity with respect to $s$ of the l.h.s.~of (6) for every continuous function $x_\tau$.

\medskip

Averaging then (6) in the above sense with respect to  $\Bbb D^m_t \{ dx_\tau \}$, dividing the result by $t-s$, and passing to the limit $s\to t$, we get a connection between the infinitesimal operators (generators) in (5) and $D^m_t$, $J^{me}_t$ and $J^m$ correspondingly, namely $$J^{me}_t =J^m -  ie    (\gamma^0    (\gamma ,A_t (x)  )- A_{0t}(x) ) \tag 7$$ which is an equality of functionals on bump functions $K^{(3)}$.

Equality (7) implies also that functional integral (5) equals the fundamental solution $D^{me}_t (x)$ of Dirac's equation
(4), or in short $$ D^{me}_t (x)= 
\Pi^t_{\tau =0} *(D^m_{ d\tau} \exp (ie  (\gamma^0    (\gamma ,A_\tau  )- A_{0\tau} )d\tau )) . \tag 8$$ Notice also that  the averaging
in  functional integral (5) goes along the set of trajectories $\{ x_\tau \}$ that lie in the support of pre-measure 
$\Bbb D^m_t \{ dx_\tau \}$, so for every such trajectory $x_\tau \in C[0,t]$ and the velocity $\dot x_\tau \in L_2 (0,t)$. Thus, since $\gamma^0 \gamma$ is the   velocity operator for  Dirac's particle (see \cite{7}), for every such trajectory
at each moment of time $\tau$ the averaging functional in (5) is $$\exp (ie \int^t_0 ((\dot x_\tau ,A_\tau )-A_{0\tau})d\tau )= \exp (ie \int^t_0 ((A_\tau ,dx_\tau )-A_{0\tau}d\tau )) \tag 9$$ hence is proportional to the identity matrix that permits to separate it from the matrix factors of the measure in (5). Notice that $$
e \int^t_0 ((A_\tau ,dx_\tau )-A_{0\tau}d\tau )=\int^t_0 Ld\tau =S^e [x_\tau ]$$ is the classical action functional for a massless particle in external field and $L$ is its Lagrangian, see \cite{5}.

\medskip

Notice also that the velocity $\dot x_\tau \in L_2 (0,t)$ of the Dirac particle
tells that during the evolution inside the light cone it does not belong to Minkowski's world (and returns there only at $\hbar \to 0$, see sect.~2 of the present article).

\medskip 

{\it Remark.} The change of variable from $\gamma^0 \gamma $ to $\dot  x_\tau$ in equation (4), where the velocity operator is present as well, cannot be performed since there 
the velocity operator acts on a function that is determined at a point, and not on a fixed trajectory with given derivative. 

\medskip 

Therefore (5) becomes $$\int  D^{me}_{ t} \varphi (x)dx = \tag 10$$ $$
\int_{\{x_\tau \}} ( \Pi_{\tau =0}^t  D^m_{d\tau} (dx_\tau ) \exp (ie  ((\dot x_\tau , A_\tau (x_\tau )) - 
 A_{0\tau}  (x_\tau )    )d\tau ))  \varphi (\int_0^t  dx_\tau )\Pi_{\tau =0}^t dx_\tau  ,  $$ or, in short, $$
D^{me}_t (x)= 
  \Pi^t_{\tau =0} * D^m_{ d\tau}  \exp (iL^e  )) ,$$ so the contribution of the external electromagnetic field is reduced to the averaging of the functional $\exp (iS)$ (where $S$ is the action of a massless charge in external electromagnetic field) with respect to the generalized functional   $\Bbb D^m_\tau \{ dx_\tau \}$ determined by the free Dirac particle of mass $m$.

\medskip 

Notice that one has $$D^m_t (x)= T^m (x)* D^{mF}_t (x)* T^m (x)$$
that relates $D^m_t (x)$ and its  Foldy-Wouthuysen presentation
$D^{mF}_t (x)$; here $T^m (x)$ is a unitary (and Hermitian) operator, see  \cite{10}, \cite{11}. So (10) implies $$D^{me}_t (x)= T^m (x)* (\Pi^t_{\tau =0} *(D^{mF}_{d\tau} \exp (iL^e)))  * T^m (x). \tag 11$$
In other words, one has exact solution of the evolution problem for the Dirac electron in exterior field using the  Foldy-Wouthuysen variables (that has diagonal structure).

We have already mentioned that the velocity of a quantum particle on a trajectory in the support of the  generalized Cauchy-Dirac functional 
$\Bbb D^m_\tau \{ dx_\tau \}$, belongs to $L_2 (0,t)$. In other words, Dirac's electron, being a relativistic quantum particle, makes spontaneous ``twitching movement" - Zittwerbewegung (see \cite{2}), but   it does not leave the light cone at that (see \cite{10}, \cite{11}). This permits to discard well-known Schr\"odinger's remark about contradictoriness of the notion of velocity (with exactly determined components) for Dirac's electron, see also \cite{10}.

\medskip

We emphasize that, as opposed to formula (1) that has mere heuristic meaning, the integral with respect to quantum measure (10) that corresponds to generalized Dirac functional $\Bbb D^m_\tau \{ dx_\tau \}$, is a rigorously defined mathematical construction.

\head 2. The correspondence principle for Dirac's electron in external electromagnetic field \endhead

Let us try to deduce the correspondence principle from the functional integral presentation (10) for the fundamental solution of Dirac's equation.

Recall (see \cite{7}) that in order to recover the dependence from Planck's constant $\hbar$ and velocity of light $c$ one should replace in the formulas $m$ by ${m_0 c}\over{\hbar}$ (here $m_0$ is the invariant mass of the electron), $e$ by $e\over{c\hbar}$, $t$ by $tc$.

\medskip

Therefore, using results of \cite{10}, we see that at $\hbar \to 0$ formula (10) yields (using $\lim_{\hbar\to 0} T^m (x)= \delta (x)$) $$ 
\int_{\{ x_\tau \}} (\Pi^t_{\tau =0} D^{mF}_{d\tau} (dx_\tau )\exp( {i\over\hbar} e      ((    dx_\tau , A_\tau (x_\tau ))-A_{0\tau}(x_\tau ))d\tau )) \varphi ( \int_0^t  dx_\tau ) \Pi_{\tau =0}^t dx_\tau , \tag 12$$ where $D^{mF}_t (x)$ is the Foldy-Wouthuysen presentation of the functional $D^m_t (x)$ (see \cite{7}), so (12) corresponds to the direct product of the functionals $$\int_{\{ x_\tau \}} (\Pi^t_{\tau =0} D^{mF}_{d\tau} (dx_\tau )\exp( {i\over\hbar} e      ((    dx_\tau , A_\tau (x_\tau ))-A_{0\tau}(x_\tau ))d\tau )) \varphi ( \ldots ,  dx_\tau ,\ldots ) \Pi_{\tau =0}^t dx_\tau . \tag 13$$

In articles \cite{10}, \cite{11} was obtained and studied the asymptotic presentation $$ \int C^m_{it}(x)|_{\hbar\to 0} \phi )x)dx =\int\exp ({i\over\hbar} m_0 l_t )\varphi (x) dx, $$ $\phi (x), \varphi (x)\in K^{(3)}$, of the elements of the matrix $$D^{mF}_t (x) =\pmatrix
C^m_{it}(x) &0&0&0\\
 0&C^m_{it}(x)&0&0\\
 0&0&\bar{C}_{it}(x)&0\\
 0&0&0&\bar{C}_{it}(x)
\endpmatrix 
$$
(here $l_t =\sqrt{t^2 - r^2}$, $r=|x|$, $m_0 l_t$ is eikanal) that became nonsingular on the light cone when we construct the asymptotic; also $
\bar{C}_{it}(x)|_{\hbar\to 0}$ disappears exponentially away  from the light cone, so $
\bar{C}_{it}(x)|_{\hbar\to 0}$ is a finite functional. It is this diagonal element in (13) gives the direct product of functionals 
$$\int_{\{ x_\tau \}} (\Pi^t_{\tau =0} \bar{C}^m_{id\tau} (dx_\tau )\exp( {i\over\hbar} e      ((    dx_\tau , A_\tau (x_\tau ))-A_{0\tau}(x_\tau )d\tau ))) \varphi ( \ldots ,  dx_\tau ,\ldots ) \Pi_{\tau =0}^t dx_\tau |_{\hbar\to 0} \tag 14$$
$$ 
=\int_{\{ x_\tau \}} \exp( - {i\over\hbar} \int^t_0 (m_0 l_\tau) -  e   ((A_\tau ,dx_\tau )-A_{0\tau}(x_\tau )d\tau )))\varphi (\ldots , dx_\tau ,\ldots ) \Pi_{\tau =0}^t dx_\tau |_{\hbar\to 0} $$
and the classical action functional of charge $-e$ in exterior electromagnetic field appears (see \cite{5}).

\medskip

This implies that at $\hbar \to 0$ the stationary phase method can be applied, so among all continuous
trajectories in the support of measure $\{ x_\tau \}$, that are used to average the functionals of (10), there arises the trajectory of minimal action, i.e., the classical trajectory of an electron in external electromagnetic field with 4-potential $\{ A_0 ,A\}$, on which the action equals the eikonal $S(x_t )$ (here $x_t$ is the position of the right end of the extremal trajectory), cf.~\cite{4}. Therefore (14) becomes $$\int\exp (-{i\over\hbar} S(x_t )) \varphi (x_t )dx_t |_{\hbar\to 0}$$

In other words, here the arguments of R.Feynman, developed by him in case of non-relativistic quantum mechanics, are valid (see \cite{3}); but, contrary to the relativistic case considered in the present article,
in Feynman's case the arguments were not rigorous due to impossibility of precise definition of Feynman's  ``path integral".

Therefore the action functional of an electron in external electromagnetic field appears in the fundamental solution of Dirac's equation only in quasiclassical approximation - see \cite{10}.

\medskip

Similar things happen at $c\to \infty$. Thus, using the standard relation between Lagrangians of  free relativistic and non-relativistic particles (see \cite{6}) and rewriting formally the direct product of functionals in (10) as the product of usual functions, we get a formal presentation of 
 the fundamental solution of   Schr\"odinger's  equation as Feynman's  ``path integral" (1).

\head Conclusion \endhead

We have derived an analog of the well-known Kac-Feynman formula for Green's function of the 
 Schr\"odinger  equation for quantum mechanics of Dirac's electron. This required an interpretation of 
Green's function of a free Dirac's electron and an electron in external electromagnetic field as functionals on bump functions.

It is important that in quantization the exponential dependence of the averaged classical action functional of
an electron in external field holds only in quasiclassical approximation.

\end